\documentclass[amsmath,amssymb,aps,superscriptaddress,twocolumn,floatfix]{revtex4-2}
\usepackage{graphicx}
\usepackage{dcolumn}
\usepackage{bm}
\usepackage[usenames]{color}
\usepackage{comment}
\usepackage{wasysym}
\usepackage{xr}

\usepackage[colorlinks=true,citecolor=blue,linkcolor=magenta]{hyperref}

\begin{document}

\title{Strange Metals from Melting Correlated Insulators in Twisted Bilayer Graphene}

\author{Peter Cha}
\affiliation{Department of Physics, Cornell University, Ithaca, New York 14853, USA}

\author{Aavishkar A. Patel}
\affiliation{Department of Physics, University of California, Berkeley, CA 94720, USA}

\author{Eun-Ah Kim}
\affiliation{Department of Physics, Cornell University, Ithaca, New York 14853, USA}

\begin{abstract}
Even as the understanding of the mechanism behind correlated insulating states in magic-angle twisted bilayer graphene converges towards various kinds of spontaneous symmetry breaking, the metallic ``normal state'' above the insulating transition temperature remains mysterious, with its excessively high entropy and linear-in-temperature resistivity.
In this work, we focus on the effects of fluctuations of the order-parameters describing correlated insulating states at integer fillings of the low-energy flat bands on charge transport. Motivated by the observation of heterogeneity in the order-parameter landscape at zero magnetic field in certain samples, we conjecture the existence of frustrating extended range interactions in an effective Ising model of the order-parameters on a triangular lattice. The competition between short-distance ferromagnetic interactions and frustrating extended range antiferromagnetic interactions leads to an emergent length scale that forms stripy mesoscale domains above the ordering transition. The gapless fluctuations of these heterogeneous configurations are found to be responsible for the linear-in-temperature resistivity as well as the enhanced low-temperature entropy. Our insights link experimentally observed linear-in-temperature resistivity and enhanced entropy to the strength of frustration, or equivalently, to the emergence of mesoscopic length scales characterizing order-parameter domains.
\end{abstract}
\maketitle

\begin{figure}
\includegraphics[width=\linewidth]{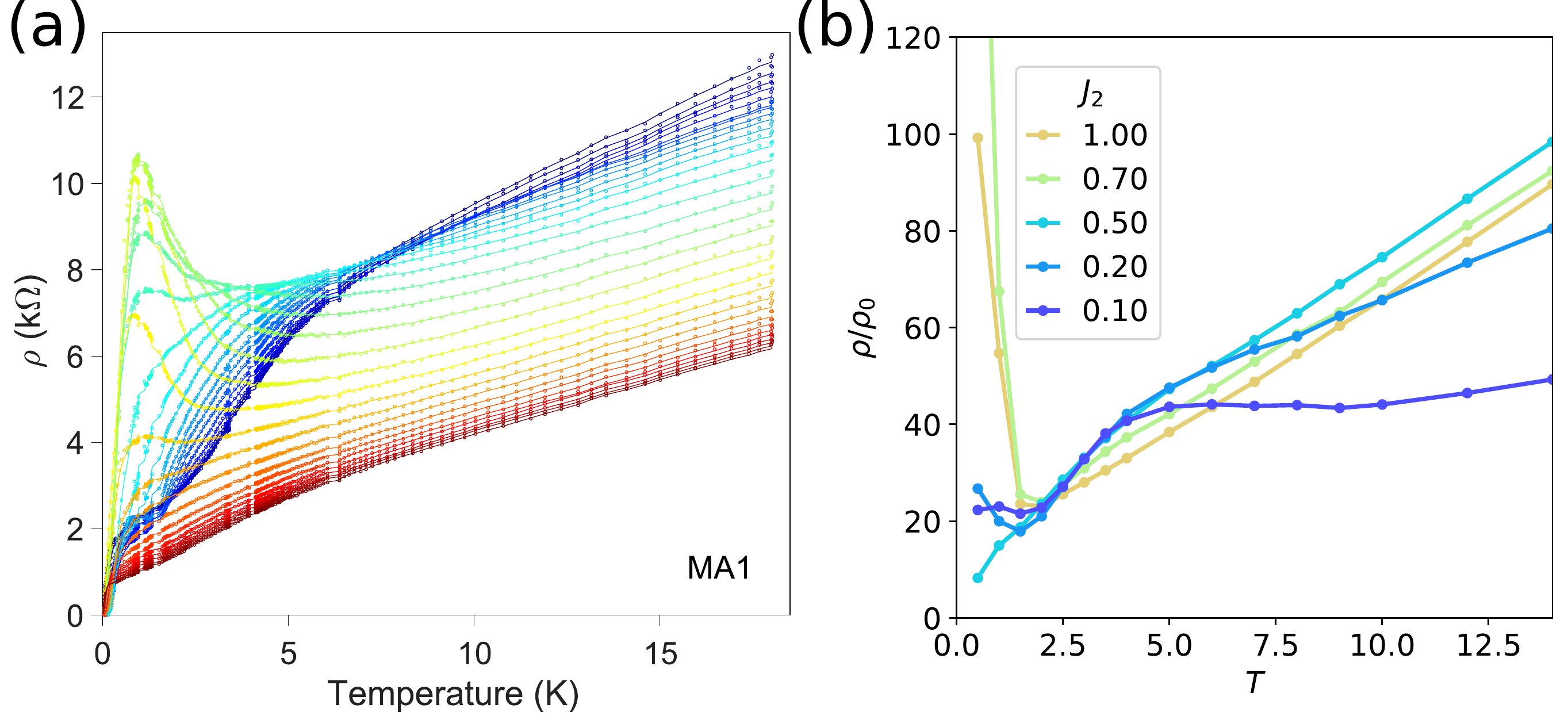}
\caption{\label{fig1} (a) DC resistivity of MATBG for various fillings from $\nu=-2.4$ (blue) to $\nu=-1.5$ (red). Plot adapted from Ref. \cite{Cao2020tlinear}.
(b) DC resistivity of the Ising effective model \eqref{model} for various values of frustrating extended-range interaction $J_2$. The vertical axis is in units of $\rho_0=\hbar/(ev_D^m)^2$, where $v_D^m$ is the effective Dirac velocity. In both (a) and (b), the progression of colors from yellow to blue indicates the lowering of the tendency to order into an insulating state at low $T$, although the parameters tuned (filling vs. Ising couplings respectively) are different.}
\end{figure}

{\it Introduction - } 
With rapid experimental developments on magic-angle twisted bilayer graphene (MATBG) reporting spin, valley, and Chern-number polarization at various integer fillings of the low-energy flat bands \cite{Sharpe2019magnetism, Lu2019corrst, Tschirhart2020imaging, Saito2020corrins, Nuckolls2020corrins}, a 
new understanding is emerging that the correlated insulating states at integer fillings \cite{Cao2018corrins, Saito2020corrins, Nuckolls2020corrins} actually arise from isospin-polarization (ISP) of gapped Dirac fermions \cite{Saito2020pomeranchuk}. On the other hand, experiments have also found that, upon heating, these insulators with ISP melt into a ``normal'' state with resistivity linear in temperature $T$ \cite{Polshyn2019tlinear, Cao2020tlinear, Ghawri2020tlinear, Pablo2021DC} (Fig. \ref{fig1}a). Further deepening the mystery is the presence of a large entropy at temperatures above the insulating transition, which is strongly enhanced non-linearly at low temperatures and rapidly quenched by an applied in-plane magnetic field (Fig. \ref{fig:entropy}a). Despite much theoretical progress on understanding the nature of correlated insulators with ISP in MATBG \cite{MacDonald2018CI,Senthil2019AHE,Bultinck2020ferr,Liu2021Nematic,Thomson2021,Bultinck2020GS,Christos2020SCCI,Lian2020tbg4}, insight into how these experimentally observed features in the normal state arise from ISP states with gapped Dirac physics remains lacking.

``Strange metal" behavior, with $T$-linear resistivity, is observed in many strongly correlated materials \cite{Martin1990, Hussey2009, Mackenzie2013} and has long remained mysterious, as such temperature dependence is inaccessible from the limit of weakly interacting quasiparticles \cite{Kivelson2016qp}. Recent studies of models with $T$-linear resistivity has shed much light on this phenomenon \cite{Song2017syk, Patel2018syk, Chowdhury2018syk, Mousatov2019hubbard, Huang2019hubbard,  Mousatov2020coldhot, Patel2019Planckian, Cha2020TL, Guo2020, Mendez2021, Altman1, Esterlis2021, GeorgesDumitrescu}. A comparative study of solvable models with local self-energy \cite{Cha2020} identified at least two distinct mechanisms towards $T$-linear resistivity at temperatures below the interaction in the ``incoherent" limit of perturbative electron hopping: a Mott-like mechanism where electronic transitions are confined to narrow bands separated by interaction scales, and a Sachdev-Ye-Kitaev-like mechanism where the electrons are quantum critical and the $T$-scaling of resistivity is a function of critical exponents. A study of a modified Hubbard model with perturbative hopping, where the ground-state degeneracy was broken by an extended-range interaction \cite{Mousatov2019hubbard, Mendez2021}, also reported $T$-linear resistivity. These recent advancements in understanding $T$-linear resistivity provide an ideal setting from which strange metal behavior in MATBG can be studied.

In this work, we capture the anticipated role of fluctuations of mesoscale domains of correlated insulator order-parameters containing ISP \cite{Tschirhart2020imaging} by modeling these fluctuations within an effective Ising model with frustration. The fluctuations affect transport through coupling to the itinerant fermions. Inspired by the observations of the (gapped) ``Dirac revival'' \cite{Zondiner2020diracrev}, we treat the itinerant fermions as Dirac fermions locally gapped by the order-parameters \cite{Zondiner2020diracrev}. We model fluctuations in local ordering and ISP tendencies using classical Monte Carlo simulations. We then calculate the temperature dependence of the resistivity and entropy and compare our findings with recent observations.

\begin{figure}
    \centering
    \includegraphics[width=\linewidth]{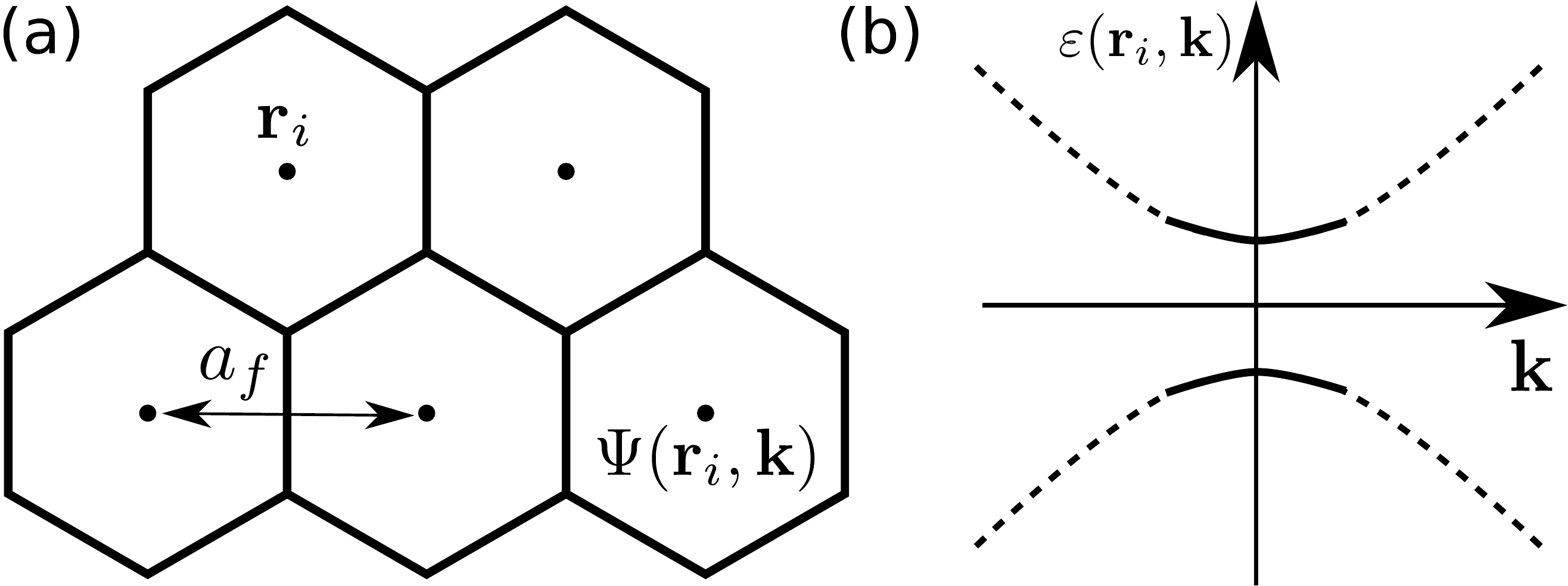}
    \caption{(a) We coarse-grain our system into hexagonal regions centered at $\mathbf{r}_i$ of size $a_f$ that form a triangular lattice. In the mean-field model \eqref{HIsing}, the Ising order-parameter $\mathcal{O}(\mathbf{r}_i)$ takes values $\pm1$ on each hexagonal plaquette. (b) Spectrum $\varepsilon(\mathbf{r}_i,\mathbf{k})$ near the Dirac points that have been gapped by interaction-induced symmetry breaking. Higher momentum fermions $\Psi(\mathbf{r}_i,\mathbf{k})$ in these regions are integrated out, leaving a mean field model for the modes near $\mathbf{k}=0$.}
    \label{fig:coarsegrain}
\end{figure}

{\it Model - } At integer fillings $\nu$, we capture the effect of the insulating order-parameters with ISP at the mean-field level as a gap-inducing potential imposed on a Dirac fermion system, {\it i.e.} a mass term, which may generically arise from integrating out long-range Coulomb interactions \cite{MacDonald2018CI,Senthil2019AHE,Bultinck2020ferr,Liu2021Nematic,Thomson2021,Bultinck2020GS,Christos2020SCCI,Lian2020tbg4}. Such mass terms can have complicated momentum dependencies in the Brillouin zone \cite{Bultinck2020GS}. However, in this work, we will keep the form of the mass terms as simple as possible and write the mass as $V(\mathbf{r})=\Psi^\dagger(\mathbf{r}) M \Psi(\mathbf{r})$. At charge neutrality, $\nu=0$, $\Psi_{\sigma\tau\eta s}(\mathbf{r})$ is a $16$-component spinor at spatial coordinate $\mathbf{r}$ and $\sigma$ indexes the $A$/$B$ sublattices, $\tau$ denotes the graphene valley ($\mathbf{K}$ or $\mathbf{K}'$), $\eta$ indexes the two mini Dirac cones of MATBG \cite{Bistritzer2011} in a valley, and $s$ denotes spin, and 
\begin{align}
M=\sum_{\alpha,\beta,\gamma,\delta=0}^3 C_{\alpha\beta\gamma\delta}~\sigma^\alpha\otimes\tau^\beta\otimes\eta^\gamma\otimes s^\delta,
\end{align}
is a $16\times16$ matrix with eigenvalues $\pm 1$. At integer fillings away from $\nu=0$, the ground state will be isospin-polarized and $\Psi$ and $M$ will have fewer components, which may include a magnetic component at certain fillings \footnote{More precisely, at filling $\nu=\pm2$, the ground state is allegedly spin polarized \cite{Senthil2019AHE,Bultinck2020GS}, so we may just drop the $s$ components. At $\nu=\pm3$, we can then drop both the $s$ and $\tau$ components \cite{Senthil2019AHE,Bultinck2020ferr}. At $\nu=\pm1$, we can write down a similar construction with $12$ component spinors.}. The mean-field order-parameter $\mathcal{O}(\mathbf{r}) = \langle V(\mathbf{r})\rangle$ can then take on the normalized values $\pm 1$, corresponding to the different degenerate mean field ground states.

\begin{figure}
\includegraphics[width=\linewidth]{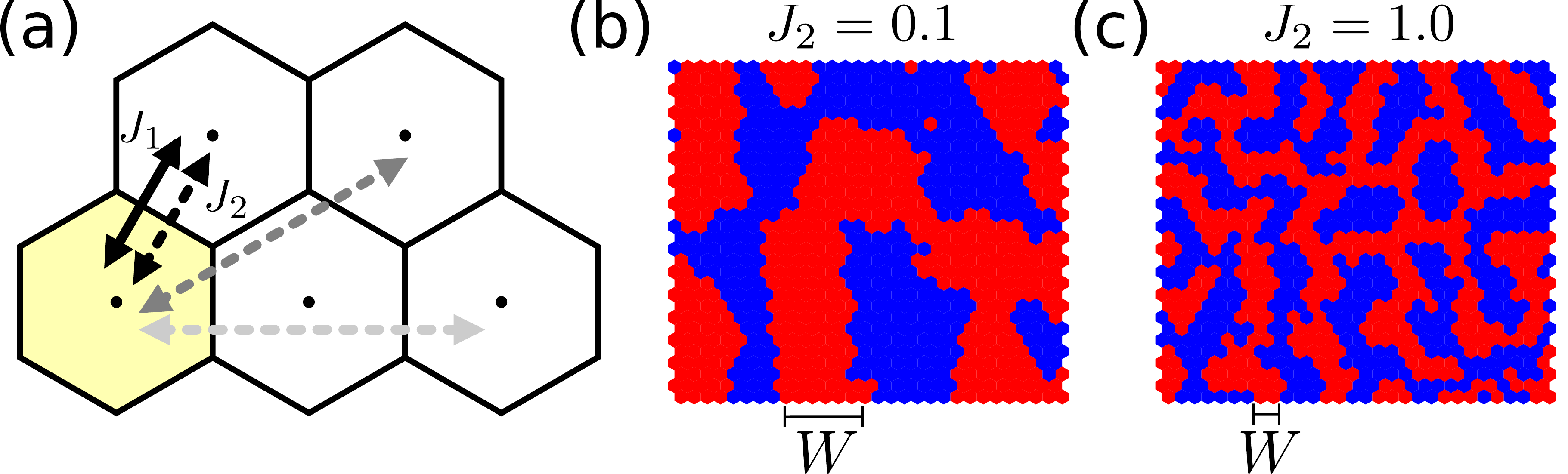}
\caption{\label{fig2} (a) A visual representation of our model \eqref{model}. Solid arrows denote nearest-neighbor ferromagnetic coupling $J_1$, and dashed arrows denote the extended-range frustrating anti-ferromagnetic interaction $J_2$, with shading indicating that the strength of the $J_2$ interaction falls exponentially in distance. (b, c) Configurations drawn from Monte Carlo sampling at $T=3.5$ and $J_2=0.1$ (b), $J_2=1.0$ (c), with the emergent length scale $W$ indicated for each configuration.
}
\end{figure}

We will then consider the effects of long wavelength fluctuations of $\mathcal{O}(\mathbf{r})$. 
To do so, we divide the system into regions centered at $\mathbf{r}_i$, which are separated by a length scale $a_f\gg a_m$, where $a_m$ is the moiré lattice constant of MATBG (Fig. \ref{fig:coarsegrain}a). Within each region, we then integrate out the higher momentum electron modes (see Supplementary Information Section I for an alternative derivation in which the fermion fields are spatially averaged) to obtain an effective Hamiltonian for the $\mathcal{O}(\mathbf{r}_i)$ (Fig. \ref{fig:coarsegrain}b). This is simply given by an Ising model of the $\mathcal{O}(\mathbf{r}_i)$, with longer than nearest-neighbor range interactions, which are in general allowed in an effective theory in which high energy fermion modes are integrated out: 
\begin{equation}
H_I = \sum_{i,j} U_{ij} \mathcal{O}(\mathbf{r}_i)\mathcal{O}(\mathbf{r}_j).
\label{HIsing}
\end{equation}
Since $M$ is larger than $2\times2$, there is actually a trivial degeneracy in the states corresponding to a given value of $\mathcal{O}(\mathbf{r}_i)$, which cancels out in the computation of correlation functions up to an overall factor of the number of flavors. Anticipating the experimental observation of mesoscopic heterogeneity \cite{Tschirhart2020imaging}, we will include in $U_{ij}$ a nearest-neighbor ferromagnetic interaction $J_1=1$ and an extended-range frustrating anti-ferromagnetic interaction $J_2$ (see Fig. \ref{fig2}a):
\begin{align}
H_I=-J_1\sum_{\langle ij\rangle}\mathcal{O}(\mathbf{r}_i)\mathcal{O}(\mathbf{r}_j)+J_2\sum_{ i,j}\mathcal{O}(\mathbf{r}_i)\mathcal{O}(\mathbf{r}_j)e^{-r_{ij}/l}.\label{model}
\end{align}
Here $r_{ij}=|\mathbf{r}_i-\mathbf{r}_j|$, where $\mathbf{r}_i$ indexes the domain sites (see Supplementary Information Section II for a comparison of $J_{1,2}$ to experimentally observed energy scales). Although the lattice geometry of $\mathbf{r}_i$ will not be important in our results, we work on a triangular lattice to preserve the $\mathcal{C}_3$ symmetry of the Moir\'e sites, and take $l=2a_f$ \footnote{This value of $l$ is sufficiently large to obtain a continuum of excited states of $H_I$ in the thermodynamic limit, which leads to a finite conductivity.} We then study the effective model of Ising variables using classical Monte Carlo simulations.

In the low-frustration ($J_2\ll 1$) and intermediate-temperature ($1<T<5$) regime, which we dub the ``stripy microemulsion'' regime \cite{Spivak2004micro}, the ensemble of spatial order-parameter configurations is dominated by stripy mesoscale domains with an emergent length scale $W$. (Fig. \ref{fig2}b) The stripy appearance of the order-parameter domains in this regime follows from a mesoscopic length-scale $W$ that satisfies $L\gg W\gg 1$, where $L$ is the system size length scale.
By contrast, away from the stripy microemulsion regime in the high-frustration ($J_2\gtrsim 1$) or in the high-temperature ($T>5$) regimes, typical configurations have $W\sim 1$, and thus each site interacts with an effectively random distribution of Ising variables in its vicinity (Fig. \ref{fig2}c).
From this observed dependence of $W$ on frustration $J_2$, we conjecture a scaling form $W/a_f=F(J_2/J_1)$ where $F(x)$ is divergent for $x\leq0$ and monotonically decreasing for $x>0$, reaching $F(x)\sim 1$ for $x\gtrsim 1$.
The emergence of $W$ can be understood from competition between the nearest-neighbor ferromagnetic interaction $J_1$, which favors long range order, with the frustrating extended range antiferromagnetic interaction $J_2$, which suppresses order. The relative strength between these interactions tunes the system between the unfrustrated Ising model, which has a divergent ordering length scale, and a frustrated emulsion-like system. At finite values of $J_2$, we find intervening configurations featuring mesoscale domains characterized by a finite length scale $W$.

{\it Transport - } The low energy current operators near the mini Dirac cones take the form \footnote{This form of the low-energy current operator is valid even when the bands are gapped and when the dispersion slows down at momenta away from the mini Dirac cones},
\begin{equation}
\mathbf{J}_{x,y}(\mathbf{r})=ev_D^m\Psi^\dagger(\mathbf{r})(\sigma^{x,y}\otimes\tau^z\otimes\eta^0\otimes s^0)\Psi(\mathbf{r}),~~(\nu=0),
\end{equation}
where $e$ is the electron charge, and $v_D^m$ is the effective Dirac velocity. In order to gap out the mini Dirac cones, the current operator must anticommute with the mass matrix $M$. It then follows that the action of $\mathbf{J}_{x}(\mathbf{r}_i)$ on the mean field ground state at $\mathbf{r}_i$ comprising of low momentum electron modes flips $\mathcal{O}(\mathbf{r}_i)\rightarrow-\mathcal{O}(\mathbf{r}_i)$. Thus, the current operator in our effective Ising model will be $\mathbf{J}_x(\mathbf{r}_i)=ev^m_D\mathcal{X}(\mathbf{r}_i)$, where $\mathcal{X}(\mathbf{r}_i)$ is the Pauli-$x$ operator that acts on the Ising variable of site $i$. We compute conductivity in the Ising effective Hamiltonian using the Kubo formula
\begin{align}
\sigma(\omega)=&\frac{i}{\hbar N}\sum_{n,m,i}P_n\frac{1-e^{-\beta E_{mn}}}{E_{mn}}\frac{\langle n|\mathbf{J}_x(\mathbf{r}_i)|m\rangle\langle m|\mathbf{J}_x(\mathbf{r}_i)|n\rangle}{\omega+i\epsilon-E_{mn}}\label{kubo},
\end{align}
where $N$ is the number of lattice sites, $n$ and $m$ index the ensemble of spatial order-parameter configurations, $P_n$ is the Boltzmann factor $P_n=e^{-\beta E_n}/Z$, and $E_{mn}\equiv E_m-E_n$ are the transition energies. In the insulating states, which correspond to the low-temperature regime of our model $T<1$,
the values of $\mathcal{O}(\mathbf{r}_i)$ are frozen. Upon heating into a metallic phase, the system may then develop fluctuations of the order-parameter that will allow current to flow between the coarse-grained order-parameter domains $\mathbf{r}_i$.

In Fig. \ref{fig1}b, we plot the DC resistivity $\rho_{DC}/\rho_0$ against temperature $T$ for various values of $J_2$, where the unit of resistivity is $\rho_0=\hbar/(ev^m_D)^2$. At $J_2=1$, which is in the high-frustration regime, we observe a slope-invariant resistivity, with no change in slope $d\rho_{DC}/dT$ from intermediate ($1<T<5$) to high ($T>5$) temperatures. By contrast, at low-frustration $0<J_2<1$ we observe two $T$-linear regimes of resistivity, at intermediate and high temperatures, with different slopes separated by a hump-like kink, indicating distinct underlying mechanisms of $T$-linear resistivity. Two distinct regimes of $T$-linear resistivity with different slopes are also seen in the experimentally measured resistivity of MATBG (Fig. \ref{fig1}a). 

Insight into the slope of $T$-linear resistivity may be gained from an investigation of the Kubo formula \eqref{kubo}. The second sum over states in \eqref{kubo} is non-vanishing only for states $m$ that differ from $n$ by a sign flip at a single domain site, and thus can be written as a sum over all sites. Then the real part of conductivity can be expressed in the form
\begin{align}
\frac{\sigma(\omega)}{\sigma_0}=\left\langle\left\langle\frac{1-e^{-\beta E_{mn}}}{E_{mn}}\frac{\epsilon}{(\omega-E_{mn})^2+\epsilon^2}\right\rangle_{m\in\text{sites}}\right\rangle_{n\sim\text{MC}},
\end{align}
where $\sigma_0=1/\rho_0$, the inner average is taken over all sites, and the outer average is taken over MC-generated configurations which follows the Boltzmann distribution. Thus after taking the limit $\epsilon\rightarrow0$ in Eq. \eqref{kubo}, we may write the conductivity as
\begin{align}
\frac{\sigma(\omega)}{\sigma_0}=\frac{1-e^{-\beta\omega}}{\omega}C(\omega),
\end{align}
where $C(\omega)=\frac{1}{N}\sum\limits_{n\sim\text{MC}}\sum\limits_{m\in\text{sites}}\delta\left(\omega-E_{mn}\right)$ is the spin-flip correlation function, satisfying $\int d\omega\ C(\omega)=1$. The DC conductivity can then be written in the simple form $\sigma_{DC}/\sigma_0=\beta C(0)$, from which we may conclude that $T$-linear resistivity is associated with a plateau of $C(0)$ in temperature. Remarkably, $C(0)$ holds simultaneous significance as the slope of $T$-linear resistivity and the MC- and site-averaged probability of an energy conserving, or soft, spin flip.

In the stripy microemulsion regime, soft spin flips may only take place on the boundaries of the stripy domains, representing fluctuations of domains that preserve $W$. Thus the soft spin flips that contribute towards $C(0)$ are the quasi-Goldstone gapless mode associated with the spontaneous translational symmetry breaking induced by the stripy heterogeneity. \footnote{More precisely, these boundary isospin fluctuations correspond to the $q=0$ translation mode of the stripy domains. Such fluctuations do not cost energy due to the isospins interacting with an equal number of up and down isospins. Since there is no preferred direction for the stripy domains, the gapless modes will be isotropic with a dispersion $\omega\sim cq$.} The spectral peak of this mode is independent of $T$ at intermediate temperature $(1<T<5)$. This $T$-independent value of $C(0)$ then sets the slope of the intermediate temperature $T$-linear resistivity seen in Fig. \ref{fig1}b. This mechanism is in contrast with the $T$-linear resistivity away from the stripy microemulsion regime, in which each Ising site fluctuates in the vicinity of an effectively random, $T$-independent distribution of Ising variables. The $J_2$ independence of the slope of resistivity in the stripy microemulsion regime is more surprising, and can be understood as a feature of the crossover between the low-frustration regime, in which the local Ising interaction is still dominant, and high-frustration regime, which is dominated by the extended-range frustration.

\begin{figure}
\includegraphics[width=\linewidth]{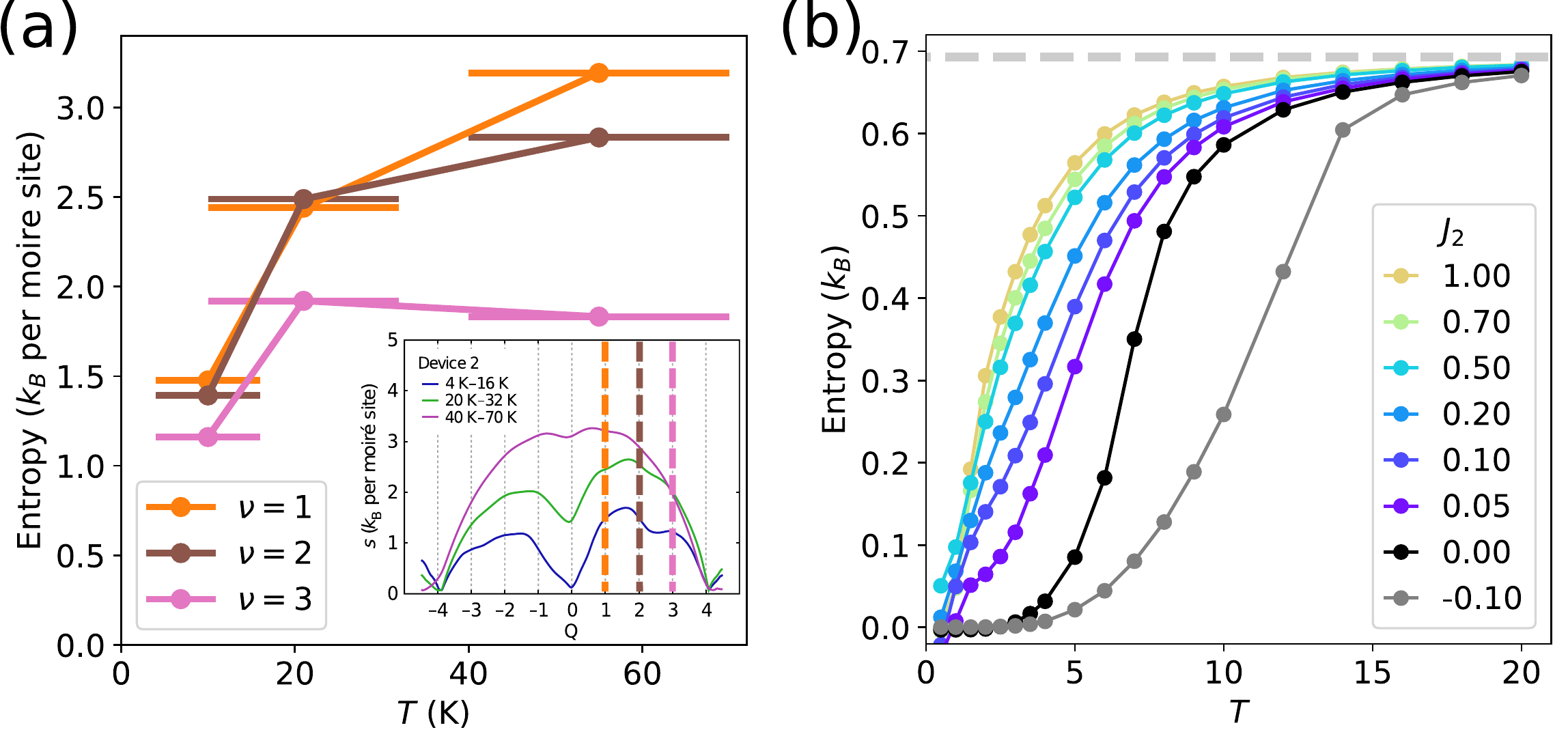}
\caption{\label{fig:entropy} (a) Entropy vs temperature $T$ at various integer fillings of MATBG adapted from Ref. \cite{Rozen2020pomeranchuk}. Inset: Original figure presented in Ref.  \cite{Rozen2020pomeranchuk}, with vertical cuts illustrating how the data was adapted for the main figure. (b) Entropy vs temperature for various values of $J_2$ in our model \eqref{model}. Dashed line at the value of $S=\ln2$ indicates the extrapolated zero-temperature residual entropy.}
\end{figure}

{\it Entropy - }Experimental investigations of integer-filled MATBG report enhanced entropy at low temperature with a non-linear dependence on temperature \cite{Rozen2020pomeranchuk} (See Fig. \ref{fig:entropy}a). We calculate entropy in the effective Ising model \eqref{HIsing} using the formula \cite{Binder1981entropy}
\begin{align}
S(\beta)=S(\beta=0)+\beta \langle E\rangle-\int_0^\beta \langle E\rangle d\beta.
\end{align}
In Fig. \ref{fig:entropy}b, we have plotted entropy vs $T$ for a range of frustration $J_2$. We observe that increasing values of $J_2$ has the effect of enhancing entropy in the low-temperature regime. The concave shape of the entropy curve at intermediate temperatures ($1<T<5$) echoes the experimental results. This enhanced entropy is readily understood within the picture of mesoscale domain formation. As the system is heated from the low-temperature limit, where the stripy domains are frozen, the Goldstone modes at the boundaries of the stripy domains are allowed to fluctuate. These fluctuating modes add an extrinsic contribution to entropy, leading to the dramatic enhancement of entropy compared to the unfrustrated Ising model (in which $J_2\leq0$) . We emphasize that the fluctuating Goldstone modes of the stripy microemulsion regime are responsible for both a sharp increase of entropy in temperature as well as $T$-linear resistivity.

We now comment on the experimentally observed sensitivity of transport and the entropy to external magnetic fields \cite{Saito2020pomeranchuk, Rozen2020pomeranchuk} observed at fillings $\nu=\pm 1,\pm 2, \pm 3$. At these fillings, since the ground states are spin polarized \cite{Senthil2019AHE,Bultinck2020ferr,Bultinck2020GS}, the spin polarization could also fluctuate like $\mathcal{O}(\mathbf{r}_i)$. Our Ising model \eqref{HIsing} may therefore indirectly couple to the Zeeman field in a model with $SU(2)$ spin:
\begin{align}
&H_{I,S} = \sum_{i,j} \left(U^0_{ij}+U^1_{ij}S^z(\mathbf{r}_i) S^z(\mathbf{r}_i)\right) \mathcal{O}(\mathbf{r}_i)\mathcal{O}(\mathbf{r}_j) \nonumber \\
&+ \sum_{ij} U^S_{ij} S^z(\mathbf{r}_i) S^z(\mathbf{r}_i) + h\sum_i S^z(\mathbf{r}_i),
\label{HIsingSpin}
\end{align}
where $h$ is the Zeeman field. In this spin-Ising coupled model, the entropy will be a sum of contributions from fluctuations in the Ising variable $\mathcal{O}(\mathbf{r}_i)$, captured by the effective Ising model \eqref{HIsing}, and fluctuations in spin $S_z$. At low temperatures, when the order-parameter sector is susceptible to mesoscale domain formation, the spin sector will also be highly susceptible to ordering. An externally applied in-plane magnetic field will then rapidly quench the entropy in the spin sector while leaving the entropy in the order-parameter sector. We speculate that such a mechanism may be responsible for the in-plane magnetic field dependence of entropy observed in Ref. \cite{Rozen2020pomeranchuk} that is difficult to explain from a standard free Dirac fermion picture.

{\it Discussion \& conclusion - }In this work, we have constructed a theory of Dirac fermions with a local mass fluctuating across space. In this theory, we have accounted for the energy cost of spatial variation in the order-parameter with an Ising variable model with nearest-neighbor ferromagnetic interaction $J_1=1$ as well a  extended-range frustrating anti-ferromagnetic interaction $J_2$, motivated by the experimentally observed absence of long range order at zero magnetic field in some MATBG samples \cite{Tschirhart2020imaging}.
By an examination of the dominant order-parameter configurations, we identify a stripy microemulsion regime at low-frustration and intermediate-temperature, characterized by an emergent mesoscopic length scale $W$ and corresponding stripy order-parameter domains.
This length scale $W$ arises from the competition between ferromagnetic order $J_1$ and frustration $J_2$, and is divergent for $J_2<0$ and monotonically decreasing for $J_2>0$, reaching $W\sim 1$ in the high-frustration regime $J_2\gtrsim 1$.
In the stripy microemulsion regime, the characteristic length scale is large $L\gg W\gg 1$, and the gapless fluctuations of the stripy mesoscale domains are responsible for both the slope of $T$-linear resistivity and the enhanced entropy. Our work therefore suggests a relationship between the slope of $T$-linear resistivity and low-temperature entropy and the appearance of mesoscale order-parameter domains in integer-filled MATBG. As a clear experimental signature of our picture, we anticipate a concurrent appearance of $T$-linear resistivity and a steep rise in entropy, both driven by fluctuating isospin domains, upon heating from the correlated insulator state.

We make a few brief comments relating our work to the existing literature. Previous works carried out fruitful studies of fluctuations from disorder in strongly correlated electronic systems \cite{liuRandomFieldDriven2016, leeColdspotsGlassyNematicity2016}. However, we emphasize that the frustration in our model is geometric rather than disorder-driven. An earlier work in the context of the cuprates considered a classical model in the strong-coupling limit and perturbative hopping \cite{Mousatov2019hubbard}. In this work, we have applied a similar strong-coupling perspective to MATBG that yielded much insight (although without the requirement of perturbative hopping). Thus the strong-coupling approach of this work is distinct from weak-coupling electron-scattering pictures of transport considered in previous theoretical studies of transport in MATBG \cite{DasSarma2019, Sharma2020}. In addition, although in this work we have identified $T>5$ as a ``high-temperature'' limit, we note that this is distinct from the ultra-high-temperature limit considered in previous studies of $T$-linear resistivity \cite{Cha2020, Perepelitsky2016hightemp}, in which $T$ is the largest scale in the problem and the thermal ensemble $e^{-\beta H}$ is proportional to the identity. Due to the extended-range of the frustrating interaction, transition energies in \eqref{kubo} can be as large as $20$, and we are therefore safely away from this regime.

Interesting directions for future work remain. An experimentally observable feature of frustration is the appearance of mesoscale order-parameter domains, reminiscent of domains that have been imaged using a superconducting quantum interference device \cite{Tschirhart2020imaging}. Our results suggest a relationship between the presence of frustration, the size of such order-parameter domains, and the observation of distinct regimes of $T$-linear resistivity and enhanced low-temperature entropy. 

It would also be interesting to consider cases where the quantization axis of $M$ is allowed to fluctuate, which would lead to vector models instead of Ising models, and to consider models where quantum fluctuations of the order parameters are also allowed (see Supplementary Information Section I for more detailed discussions). 

Finally, while we have focused on the integer fillings in this work, a different kind of $T$-linear resistivity, this time at low temperatures below the ordering temperature ($\lesssim 10 \mathrm{K}$) of the correlated insulator phase at integer fillings, is also observed when the system is doped away from integer fillings, with a magnitude much smaller than $h/e^2$ per Dirac flavor \cite{Polshyn2019tlinear, Cao2020tlinear, Ghawri2020tlinear, Pablo2021DC}. This $T$-linear resistivity also displays the phenomenon of ``Planckian dissipation" \cite{Cao2020tlinear,Pablo2021DC}, which likely originates from a Fermi surface \cite{DasSarma2019, Sharma2020, Altman1, Esterlis2021}. It would therefore be interesting to explore in detail how the resistivity evolves as a function of $T$ away from integer fillings. We expect the parameters $J_{1,2}$ in (\ref{HIsing}) to smoothly deform as we move away from integer fillings and the fluctuations of $\mathcal{O}(\mathbf{r})$ to therefore still persist (see Supplementary Information Section I for a more detailed discussion of fractional fillings).
The low temperature behavior of the resistivity at fractional fillings must be different from that at integer fillings since the ground states of fractional and integer fillings are rather different, {\it i.e.}, a Fermi surface instead of an insulator. Nevertheless, the Fermi surface will be washed out at higher temperatures due to strong quasiparticle scattering, suppressing its contribution to the conductivity. Thus, we expect that the fluctuations of the order parameter $\mathcal{O}(\mathbf{r})$ considered in our model will then contribute significantly to the conductivity. This might help explain why a similar slope of the $T$-linear resistivity is experimentally observed at $T\gtrsim 10 \mathrm{ K}$ independent of the filling \cite{Cao2020tlinear,Pablo2021DC}. Indeed, varying $J_2/J_1$ in our model leads to a qualitatively similar variation of $\rho(T)$ as varying the filling in experiments, at intermediate and high temperatures (Fig. \ref{fig1}).

{\it Acknowledgements - } P.C. and E.-A.K. were supported by NSF award OAC-1934714. A.A.P. acknowledges support from the Miller Institute for Basic Research in Science.

\bibliographystyle{apsrev4-2}

\bibliography{tbg}

\end{document}